\documentclass[11pt, letter]{article}

\usepackage[utf8]{inputenc}
\usepackage[final]{microtype}
\usepackage{authblk}
\usepackage[letterpaper, margin=1in]{geometry}

\usepackage[hyphens]{url}
\usepackage[square,numbers]{natbib}
\usepackage[hidelinks]{hyperref}
\usepackage{breakurl}

\usepackage[inline]{enumitem}

\usepackage{caption}
\usepackage{subcaption}
\usepackage[frozencache,cachedir=.]{minted}

\usepackage{graphicx}
\usepackage{amsmath,amsfonts,amssymb,amsthm}
\theoremstyle{definition}

\usepackage{listings}
\usepackage[ruled,vlined,linesnumbered]{algorithm2e}

\usepackage{mathtools}

\let\ket\relax
\DeclarePairedDelimiter{\ket}{\lvert}{\rangle}
\let\bra\relax
\DeclarePairedDelimiter{\bra}{\langle}{\rvert}
\let\braket\relax
\DeclarePairedDelimiterX{\braket}[2]{\langle}{\rangle}{#1 \delimsize\vert #2}
\DeclarePairedDelimiter\ceil{\lceil}{\rceil}

\title{Using Quantum Computers to Speed Up Dynamic~Testing~of~Software}

\begin{document}
\author[ ]{Andriy Miranskyy}

\affil[ ]{Department of Computer Science, Toronto Metropolitan University (formerly~Ryerson~University), Toronto, ON M5B2K3 Canada}
\affil[ ]{\textit{avm@ryerson.ca}}
\date{}
\maketitle

\begin{abstract}

Software under test can be analyzed dynamically, while it is being executed, to find defects. However, as the number and possible values of input parameters increase, the cost of dynamic testing rises.

This paper examines whether quantum computers (QCs) can help speed up the dynamic testing of programs written for classical computers (CCs).
To accomplish this, an approach is devised involving the following three steps: 
\begin{enumerate*}[label={(\arabic*)}]
    \item converting a classical program to a quantum program;
    \item computing the number of inputs causing errors, denoted by $K$, using a quantum counting algorithm; and
    \item obtaining the actual values of these inputs using Grover’s search algorithm.
\end{enumerate*}

This approach can accelerate exhaustive and non-exhaustive dynamic testing techniques. On the CC, the computational complexity of these techniques is $O(N)$, where $N$ represents the count of combinations of input parameter values passed to the software under test. In contrast, on the QC, the complexity is $O(\varepsilon^{-1} \sqrt{N/K})$, where $\varepsilon$ is a relative error of measuring $K$.

The paper illustrates how the approach can be applied and discusses its limitations. Moreover, it provides a toy example executed on a simulator and an actual QC.
This paper may be of interest to academics and practitioners as the approach presented in the paper may serve as a starting point for exploring the use of QC for dynamic testing of CC code.

\end{abstract}

\maketitle

\section{Introduction}\label{sec:intro}
Dynamic testing is an essential part of the software development process. It helps to uncover and eliminate the software's defects by executing the software and comparing actual and expected outputs.

The ideal scenario would be to test all possible execution paths in a software program using all possible inputs in order to detect defects. This approach is referred to as \emph{exhaustive testing}. While such testing will not eliminate all defects, it will improve our chances of finding less obvious failure-inducing bugs (which is crucial for mission-critical software~\cite{hynninen2018software}). For a sufficiently large program, this approach is not feasible on a classical computer (CC), as the number of execution paths grows exponentially. 

To reduce the number of paths we must execute, we use various testing techniques (e.g., equivalence partitioning, boundary value analysis, and orthogonal array testing~\cite{pressman2005software}). We will refer to these techniques collectively as \emph{non-exhaustive testing}. Although these techniques can help us detect some defects in a reasonable amount of time, they may miss some defects that would be found through exhaustive testing. 

The \emph{contribution} of this paper is to demonstrate that a quantum computer (QC) can accelerate dynamic testing\footnote{It was shown that various phases of the software development lifecycle, including testing, can be accelerated by QCs~\cite{miranskyy2022quantum}. However, testing of execution paths was not addressed in~\cite{miranskyy2022quantum}.}, both exhaustive and non-exhaustive. We will explore how Grover's algorithm~\cite{grover1996fast, boyer1998tight} combined with a quantum counting algorithm~\cite{boyer1998tight, brassard1998quantum, aaronson2020quantum} can help us accomplish this goal.

The rest of the paper is organized as follows. Section~\ref{sec:framework} introduces quantum algorithms required to find defects in CC software. Section~\ref{sec:testing}  provides an overview of dynamic testing on a QC and illustrates it with examples. Section~\ref{sec:limitations} discusses the limitations of this testing approach. Finally, Section~\ref{sec:summary} concludes the paper.

\section{Required algorithms and tools}\label{sec:framework}
Let us explore algorithms and tools that enable dynamic software testing using QC.

\subsection{Grover's algorithm}\label{sec:grover}
Grover's algorithm~\cite{grover1996fast} is often described as a quantum algorithm for searching an unsorted database with $N$ entries for a single item. An unsorted database search on a CC requires $O(N)$ computations. Grover's algorithm requires only $O(\sqrt{N})$ computations on a QC; it is known that Grover's algorithm is optimal up to a sub-constant factor~\cite{bennett1997strengths}. This algorithm can be generalized to search for $K$ items requiring $O(\sqrt{N/K})$ calculations~\cite{boyer1998tight}.

However, Grover's algorithm is designed to solve a more generic problem. Instead of searching for an index of an element in a database, it finds (with a high probability) an input to function $f$ that produces a particular output value.

Grover's algorithm requires a specific form for function $f$. It is a boolean function with bit string input $x = \{0, 1\}^n$ and a return value of $0$ or $1$. It is important to note that input $x$ can be interpreted as an integer (by mapping the bit string to an integer): $x = \{0, 1, \ldots, N-1\}$, where $N = 2^n$.

By convention, the search criterion is satisfied when $f(x) = 1$. We are interested in finding specific values of $x$ that yield $f(x) = 1$. We denote these values as $\omega$. We can think of the search as an inversion of $f(x) = 1$, i.e., $\omega = f^{-1}(1)$.

Then, we need to define a subroutine $U_\omega$ (often called an oracle) that acts as follows:
\begin{equation*}
    U_\omega \ket{x} =
    \begin{cases}
      \ket{x}   & \textrm{if}~ x \neq \omega \textrm{, i.e., } f(x) = 0,\\
    - \ket{x} & \textrm{if}~ x = \omega \textrm{, i.e., } f(x) = 1,
    \end{cases}
\end{equation*}
where $U_\omega$ is a diagonal unitary matrix:
\begin{equation*}
    U_\omega = (-1)^{f(x)} = 
    \begin{bmatrix}
    (-1)^{f(0)} & 0 & \cdots & 0 \\
    0 & (-1)^{f(1)} & \cdots & 0 \\
    \vdots & 0  & \ddots & \vdots \\
    0 & 0 & \cdots & (-1)^{f(N - 1)}
    \end{bmatrix}.
\end{equation*}
To represent the oracle on a QC, we need only $n = \ceil{\log_2 N}$ qubits, where $\ceil{\cdot}$ is  the ceiling function. From here on, for simplicity's sake, we will refer to $f(x)$, rather than $U_\omega$, as an \emph{oracle}.

Algorithm 1 shows the steps in Grover's algorithm once the $U_\omega$ is defined. Grover's iteration count $r(N,K)$ is given by
\begin{equation*}\label{eq:iter_cnt}
    r(N,K) = \ceil[\Bigg]{\frac{\pi}{4} \sqrt{\frac{N}{K}}},
\end{equation*}
where $K$ is the number of $\omega$ values yielding $f(x) = 1$. The computation of $r(N,K)$ is trivial if $K$ is known. The next step is to figure out how to calculate $K$ without knowing it beforehand.

\begin{algorithm}[ht]
    \caption{Grover's algorithm.}\label{alg:grover}
    Initialize the QC to the uniform superposition over all states: $\ket{s} = \frac{1}{\sqrt{N}}\sum_{x=0}^{N-1}{\ket{x}}$;  \\
    \For{$i \gets 1$ \KwTo $r(N, K)$}{ 
        \tcc{Compute "Grover iteration"}
        Apply $U_\omega$; \\
        Apply the Grover diffusion operator $U_s = 2 \ket{s}\bra{s} - I$, where $I$ denotes an identity matrix;
    }
    Measure the resulting quantum state.
\end{algorithm}

\subsection{Quantum counting algorithm}\label{sec:quantum_counting}
If the value of $K$ is unknown, it can be estimated on a QC to within relative error $\varepsilon$ using the quantum counting algorithm~\cite{boyer1998tight, brassard1998quantum}, which combines ideas from Grover's and Shor's~\cite{shor1994algorithms} algorithms. Quantum counting has a computational complexity of $O(\varepsilon^{-1} \sqrt{N/K})$.
Quantum counting takes the oracle as input. Another user-supplied input, which controls output quality, is $\varepsilon$. The algorithm returns a value of $K$, which is  $\le N$.

There exist several simplified versions of the counting algorithm (see~\cite{aaronson2020quantum} for a review). For example, \cite{aaronson2020quantum} improved the algorithm by removing Shor's-specific parts.
Computational complexity of \cite{aaronson2020quantum} changes to $O(\varepsilon^{-1} \log(\delta^{-1}) \sqrt{N/K})$, where $\delta$ is another input parameter\footnote{Algorithm~\cite{aaronson2020quantum} returns $\hat K$, which ``satisfies $K (1-\varepsilon) < \hat K < K (1 + \varepsilon)$ with probability at least $1-\delta$''~\cite{aaronson2020quantum}.}. 

\subsection{Reduce software program to an $f(x)$}\label{sec:cc_conversion}
Now let us take a look at converting CC software into QC software. Assume that a classical function $g(\vec{z})$ represents an arbitrary CC software program. In order to integrate $g(\vec{z})$ into Grover's algorithm, we must reduce it to $f(x)$. Below, we will discuss how to alter the $g(\vec{z})$ body, inputs, and outputs.

\subsubsection{Body of $g(\vec{z})$}\label{sec:cc_conversion_body}
It is theoretically possible to convert any CC program into a QC program (because QC implements a quantum Turing machine~\cite{benioff1980computer, aaronson2005quantum}), but there are limitations and practical considerations (see Section~\ref{sec:limitations}). Essentially, a classical algorithm is converted into reversible circuits~\cite{miller2003transformation}, which are then converted into quantum circuits~\cite{bojic2014approach}. These conversions can be done automatically.  

Several nascent tools for converting classical to quantum circuits already exist. Qiskit, for example, allows compiling a classical function to a quantum circuit transparently (by adding the decorator \texttt{@classical\_function}). Qiskit converts classical circuits to quantum circuits using the Tweedledum library~\cite{schmitt2022tweedledum}. Currently, the compilation is limited to binary inputs/outputs and logical operators (see Section~\ref{sec:example} for an example). Another example is the staq toolkit, which can convert classical Verilog circuits to a quantum assembly language (QASM)~\cite{amy2020staq}. Finally, StaffCC compiler~\cite{javadiabhari2014scaffcc} extension RKQC~\cite{holmes2016rkqc} is also capable of compiling classical circuits to QASM (including integer addition and multiplication operations~\cite{wang2016improved}).

Hopefully, the abovementioned tools will evolve in the future to enable the conversion of any classical program.

\subsubsection{Input to $g(\vec{z})$} 
Section~\ref{sec:grover} discussed the possibility of interpreting input $x$ into $f$ as a bit string. Furthermore, bit strings can represent arbitrary complex data structures. Thus, to convert $\vec{z}$ into $x$, we must convert $\vec{z}$ into a bit string.

For example, imagine that $\vec{z}$ has two elements: an integer $z_1$ and a 3-character string $z_2$, i.e., $g(\vec{z}) = g(z_1, z_2)$. Furthermore, suppose that storing an integer and a string character requires $64$ and $8$ bits, respectively. We will need $88 = 64 + 3 \times 8$ bits to represent the input, i.e., $N = 2^{88}$. 

\subsubsection{Output of $g(\vec{z})$} 
Software programs may produce any output (or no output at all), whereas $f(x)$ should return either zero or one. We can satisfy this requirement by wrapping\footnote{We can also leverage try-catch blocks in high-level programming languages.} $g(\vec{z})$ in code that returns an exit code of $0$ on success and $1$ on error (similar to operating system exit status codes).

\section{Dynamic testing}\label{sec:testing}
Algorithm~\ref{alg:bug_finder} shows how QC can be used to test a software product dynamically. The algorithm leverages the ``building blocks'' covered in Section~\ref{sec:framework}. We will discuss two cases of how to use the algorithm for exhaustive and non-exhaustive testing below.

\begin{algorithm}[t]
\caption{Using a QC to dynamically test software for a CC. Only Steps~\ref{alg:bug_finder_step2} and~\ref{alg:bug_finder_step3} are computed (mainly) on a QC; the rest are executed on a CC.}\label{alg:bug_finder}
    \SetKwInOut{Input}{Input}
    \SetKwInOut{Output}{Output}
    \Input{CC software $g(\vec{z})$} 
    \Output{A list of values of $\vec{z}$ causing $g(\vec{z})$ to return errors denoted by $v$}
    Convert CC software $g(\vec{z})$ to $f(x)$ as discussed in Sec.~\ref{sec:cc_conversion}; \\\label{alg:bug_finder_step1}
    Compute $K$ using a quantum counting algorithm described in Sec.~\ref{sec:quantum_counting}; \\\label{alg:bug_finder_step2}
    \uIf{K = 0}{
        $v \gets$ empty list \tcp*{all tests pass}
    }\uElseIf{K = N}{
        $v \gets$ all possible values of $\vec{z}$ \tcp*{all tests fail}
    }\Else{\label{alg:bug_finder_if} \tcc{some tests fail, i.e., $0 < K < N$}
        Find all $\omega$ (values of $x$ causing bugs, i.e., those where $f(x)=1$) using Grover's algorithm shown in Sec.~\ref{sec:grover}; \\\label{alg:bug_finder_step3}
        $v \gets$ convert $\omega$ to $\vec{z}$ format (see example in Eq.~\ref{eq:conversion}). \\ \label{alg:bug_finder_step4}
    }
\end{algorithm}

\subsection{Exhaustive testing}
Suppose we need to implement a function $g(\vec{z})$. We want to ensure that none of the possible values of $\vec{z}$ result in an error in our implementation of this function. We will follow the steps described in Algorithm~\ref{alg:bug_finder} to accomplish this. Here is a toy example of using the algorithm.

\subsubsection{Example}\label{sec:example}
Suppose we implement a function $g(z_1, z_2)$, where $z_1$ is a 2-bit integer and $z_2$ is a boolean variable. The next step is to test our implementation of $g(z_1, z_2)$. 

Suppose we inject two defects, e.g., $g(z_1, z_2)$ returns incorrect results when $z_1 = 3$ and $z_2 = \textrm{false}$ or when $z_1 = 2$ and $z_2 = \textrm{true}$. Algorithm~\ref{alg:buggy_code} gives a trivial example of such code. Code review of Algorithm~\ref{alg:buggy_code} immediately reveals the problem. Of course, in real-world scenarios, we would not know these values of $\vec{z}$ upfront; this simplification is for demonstration purposes only.

\begin{algorithm}[t]
    \SetKwInOut{Input}{Input}
    \SetKwInOut{Output}{Output}
    \Input{$z_1$, $z_2$} 
    \Output{$y$}
    \eIf{ $(z_1 = 3$ and $z_2 = \textrm{false})$ or $(z_1 = 2$ and $z_2 = \textrm{true})$ }{
        \tcc{Error-prone business logic goes here}
        $y \gets 1$
    }{
        \tcc{Correct business logic goes here}
        $y \gets 0$
    }
    \caption{Sample CC pseudocode under test.}
    \label{alg:buggy_code}
\end{algorithm}

Step~\ref{alg:bug_finder_step1} of Algorithm~\ref{alg:bug_finder} converts $g(z_1, z_2)$ into $f(x)$. As $z_1$ and $z_2$ can be represented by 3-bit strings, $N = 2^3 = 8$. The first two bits of the bit string store an integer $z_1$, while the last bit stores a boolean variable $z_2$.

We will implement our example using Qiskit; code listing can be found on Zenodo~\cite{andriy_miranskyy_2022_7065888}.

\begin{figure}[ht]
    \centering
    \begin{subfigure}[b]{0.5\columnwidth}
        \centering
        \includegraphics[width = 0.4\textwidth]{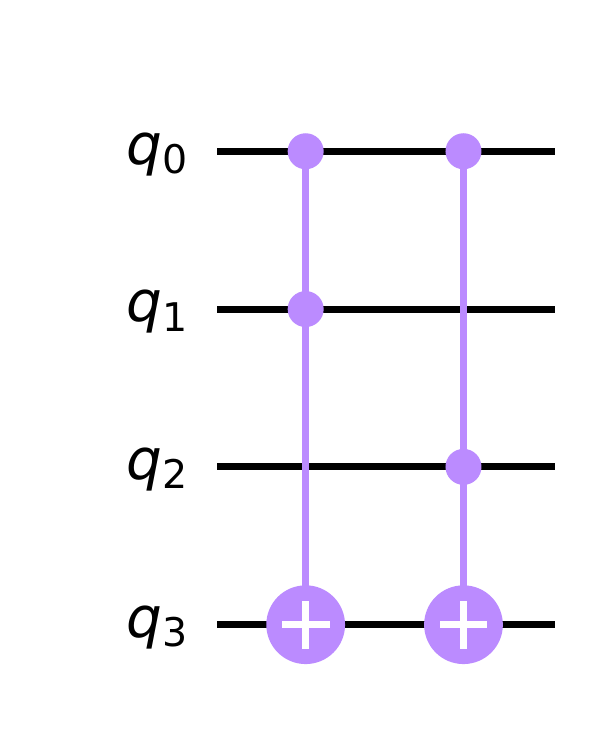}
        \caption{Circuit}
    \end{subfigure}
    \hfill
    \begin{subfigure}[b]{0.49\columnwidth}
        \centering
        \begin{minted}[fontsize = \footnotesize, numbersep = 2mm, linenos = true, autogobble]{cpp}
            OPENQASM 2.0;
            include "qelib1.inc";
            qreg q[4];
            ccx q[0],q[1],q[3];
            ccx q[0],q[2],q[3];
        \end{minted}
        \caption{Assembly code}
    \end{subfigure}
    \caption{
    The quantum circuit obtained by converting the bit-flip oracle in Listing~\ref{lst:f_x}. In OpenQASM, \texttt{ccx} denotes controlled-controlled-not (Toffoli) gate. $q_0$ corresponds to high bit of $z_1$, $q_1$ --- to low bit of $z_1$, and $q_2$ --- to $z_2$. Oracle's return value is stored in $q_3$.} 
    \label{fig:bit_flip_oracle}
\end{figure}

As was described in Section~\ref{sec:cc_conversion}, converters from a classical to a quantum circuit are still in their infancy, and Qiskit operates only on binary variables. Thus, we must manually reduce the code in Algorithm~\ref{alg:buggy_code} to the format understood by Qiskit (eventually, converters and compilers will do this automatically). Listing~\ref{lst:f_x} shows the result of the conversion to $f(x)$. Once in this form, Qiskit’s \texttt{@classical\_function} decorator automatically converts $f(x)$ to the quantum circuit (so-called bit-flip oracle~\cite{nielsen_chuang_2010}) shown in Figure~\ref{fig:bit_flip_oracle}.

\begin{listing}[ht]
\begin{minted}[fontsize = \footnotesize, numbersep = 2mm, linenos = true, autogobble]{python}
@classical_function
# Int1 is Qiskit's bit-level type
def oracle(z1_high_bit: Int1, z1_low_bit: Int1, z2: Int1) -> Int1:
    return (z1_high_bit and z1_low_bit and not z2) or \
           (z1_high_bit and not z1_low_bit and z2)
\end{minted}
\caption{A bit-flip oracle obtained by converting $g(z_1, z_2)$ to $f(x)$.}
\label{lst:f_x}
\end{listing}

\begin{figure}[ht]
    \centering
    \begin{subfigure}[b]{0.5\columnwidth}
        \centering
        \includegraphics[width = 0.75\textwidth]{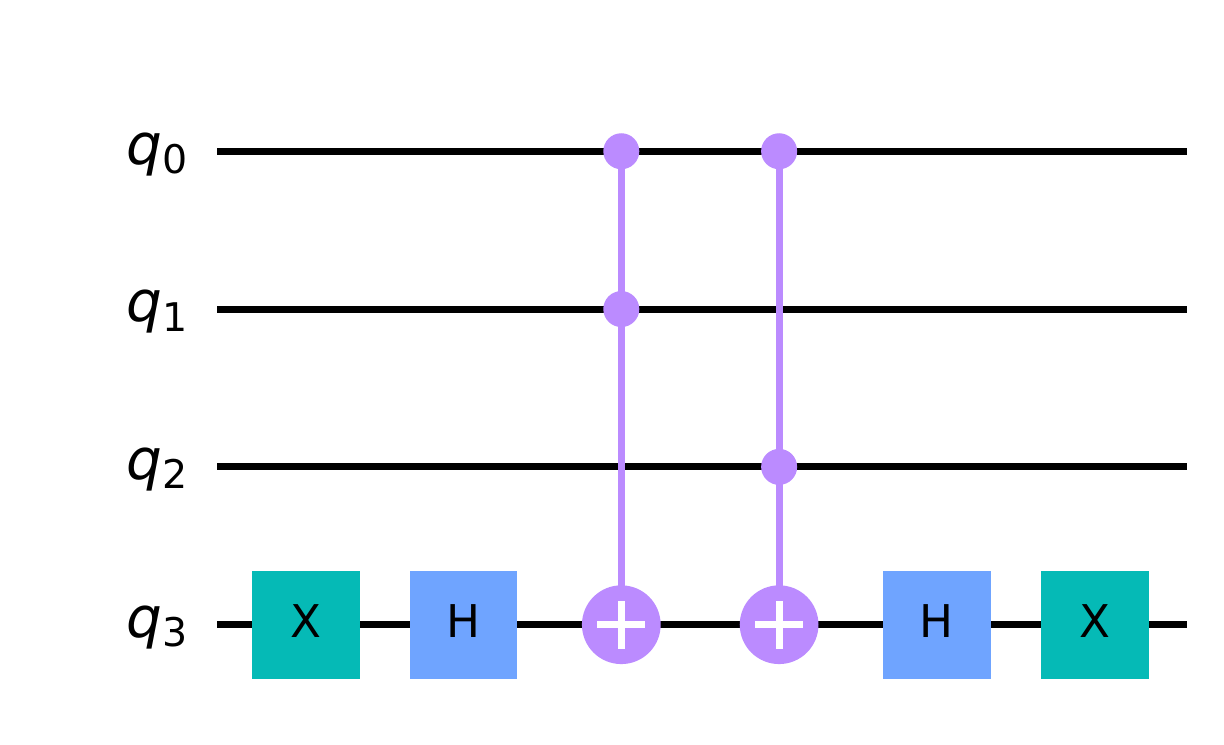}
        \caption{Circuit}
    \end{subfigure}
    \hfill
    \begin{subfigure}[b]{0.49\columnwidth}
        \centering
        \begin{minted}[fontsize = \footnotesize, numbersep = 2mm, linenos = true, autogobble]{cpp}
            OPENQASM 2.0;
            include "qelib1.inc";
            qreg q[4];
            x q[3];
            h q[3];
            ccx q[0],q[1],q[3];
            ccx q[0],q[2],q[3];
            h q[3];
            x q[3];
        \end{minted}
        \caption{Assembly code}
    \end{subfigure}
    \caption{
    The quantum circuit obtained by converting the bit-flip oracle in Figure~\ref{fig:bit_flip_oracle} to phase-flip oracle. In OpenQASM, \texttt{x} denotes NOT gate, \texttt{h} -- Hadamard gate.}
    \label{fig:phase_flip_oracle}
\end{figure}

Using Grover's algorithm requires a phase-flip oracle~\cite{nielsen_chuang_2010}. Bit-flip oracles are converted to phase-flip oracles by placing their return values between NOT and Hadamard gates as shown in Listing~\ref{lst:phase_oracle}. The phase-flip oracle circuit, which we pass to the search algorithm, is shown in Figure~\ref{fig:phase_flip_oracle}.

\begin{listing}[ht]
\begin{minted}[fontsize = \footnotesize, numbersep = 2mm, linenos = true, autogobble]{python}
def get_phase_flip_oracle(bitflip_oracle, n):
    phase_oracle = QuantumCircuit(n+1)
    phase_oracle.x(n)
    phase_oracle.h(n)
    phase_oracle.compose(bitflip_oracle.synth(), inplace=True)
    phase_oracle.h(n)
    phase_oracle.x(n)
    return phase_oracle
input_qubits_count = 3  
phase_flip_oracle = get_phase_flip_oracle(oracle, input_qubits_count)
\end{minted}
\caption{Conversion of the bit-flip oracle from Listing~\ref{lst:f_x} to a phase-flip oracle.}
\label{lst:phase_oracle}
\end{listing}

In Step~\ref{alg:bug_finder_step2} of Algorithm~\ref{alg:bug_finder}, we pass the oracle's circuit to the quantum counting algorithm\footnote{Qiskit's implementation of the quantum counting algorithm~\cite{brassard1998quantum} is illustrated in~\cite[Sec. 3.9]{Qiskit-Textbook}.}, which returns $K$, the number of inputs causing defects. In our case, $K=2$. As $K \neq 0$ and $K \neq N$, we go to Step~\ref{alg:bug_finder_step3}.

Step~\ref{alg:bug_finder_step3} of Algorithm~\ref{alg:bug_finder} involves passing oracle's circuit to Grover's algorithm to find specific input values that cause failures. Using the $K$ value from Step~\ref{alg:bug_finder_step2}, we can calculate the number of iterations of Grovers' algorithm: $r(N, K) = r(8, 2) = 2$. 

Execution is performed on the Qiskit v.0.37.0 Aer simulator v.0.10.4 (providing an ideal noise-free QC) as well as on the IBM Quantum r4T Processor v.1.0.38 with five qubits, namely, ibmq\_lima \cite{ibm_quantum}. With ibmq\_lima, we compare the output based on Qiskit's simulator of ibmq\_lima as well as the output from the actual QC.

\begin{figure*}[ht]
    \centering
    \begin{subfigure}[b]{0.32\textwidth}
        \centering
        \includegraphics[width = \textwidth]{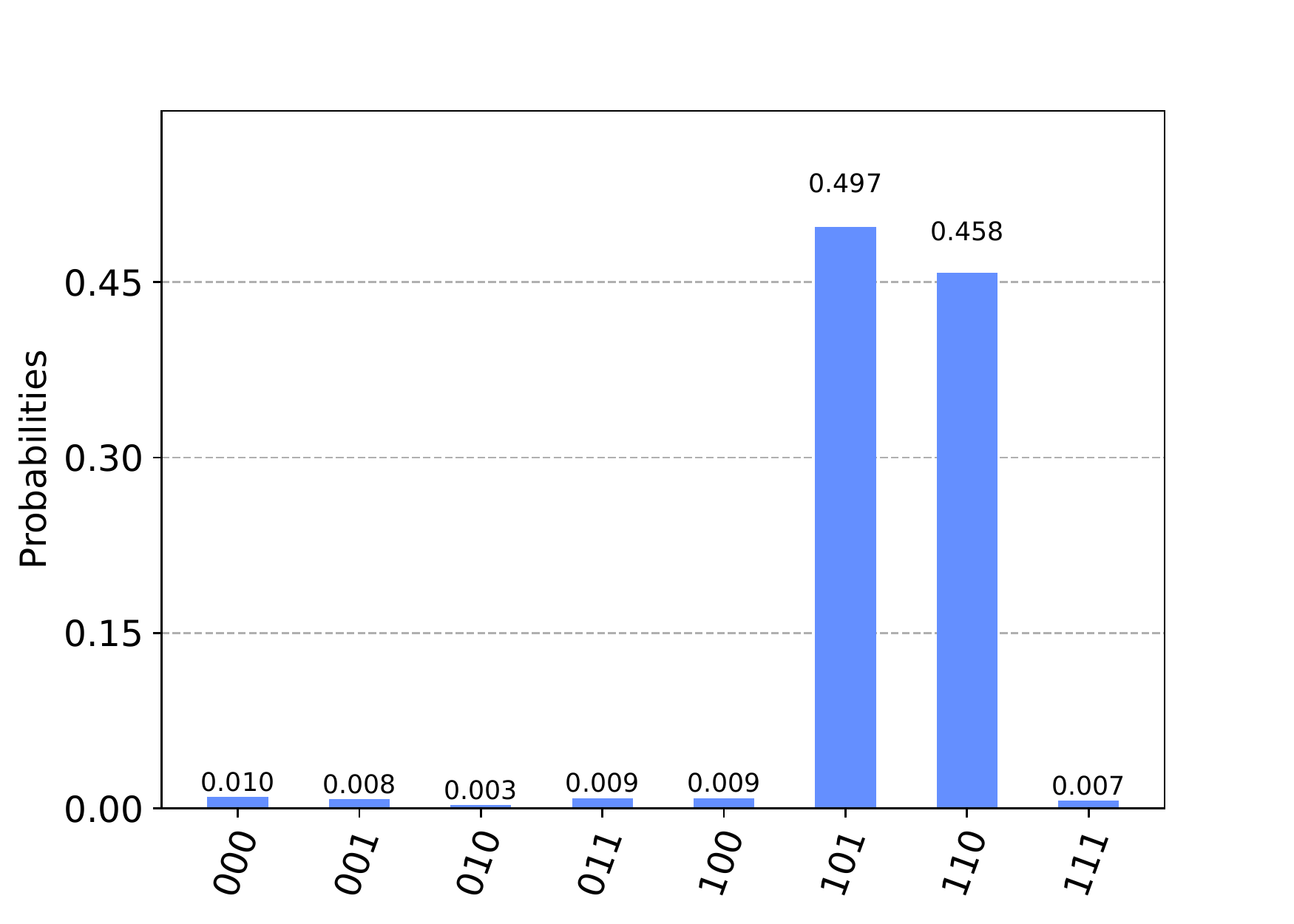}
        \caption{Qiskit Aer simulator}
        \label{fig:sim}
    \end{subfigure}
    \begin{subfigure}[b]{0.32\textwidth}
        \centering
        \includegraphics[width = \textwidth]{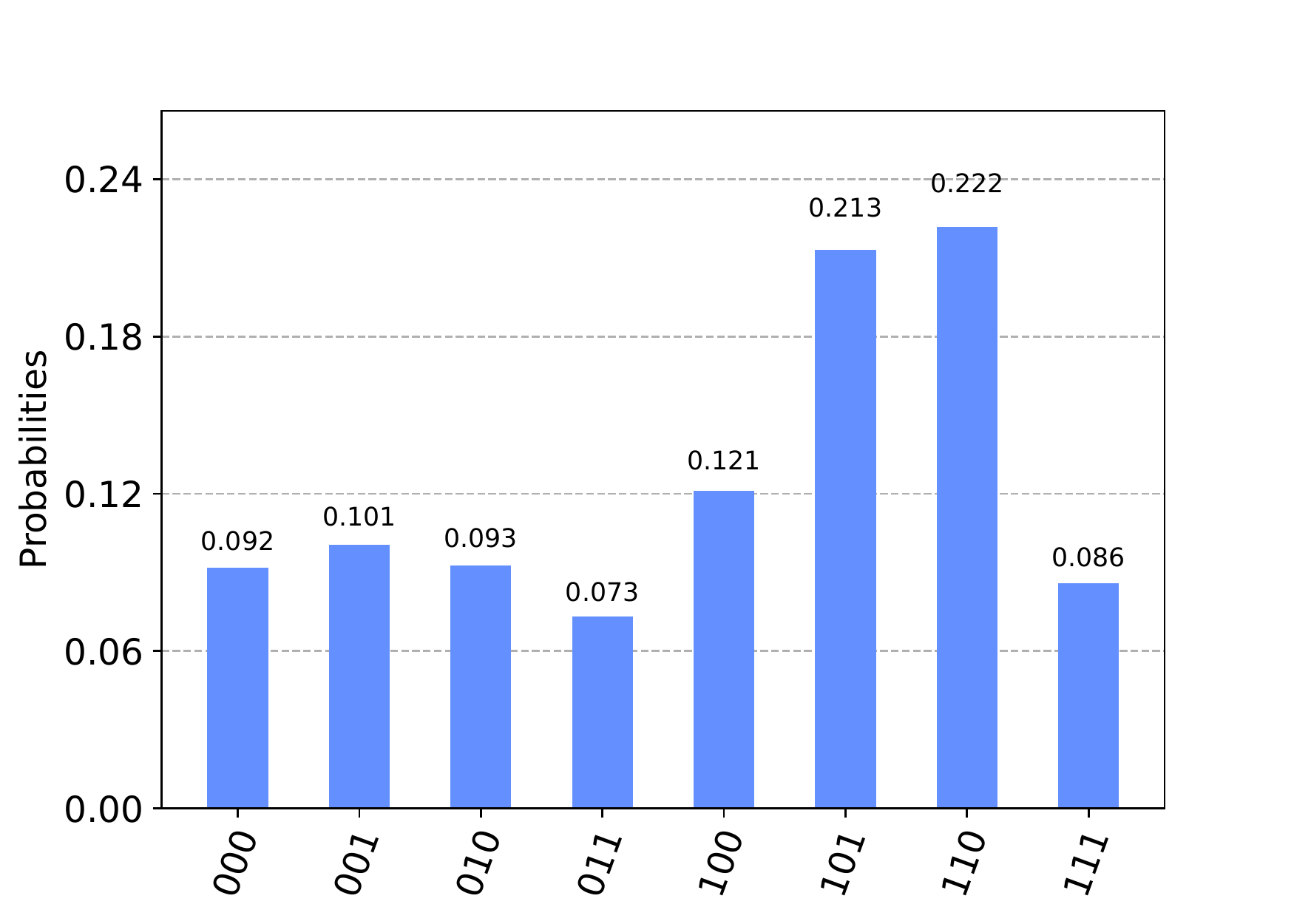}
        \caption{Qiskit simulator of ibmq\_lima}
        \label{fig:lima_sim}
    \end{subfigure}
    \begin{subfigure}[b]{0.32\textwidth}
        \centering
        \includegraphics[width = \textwidth]{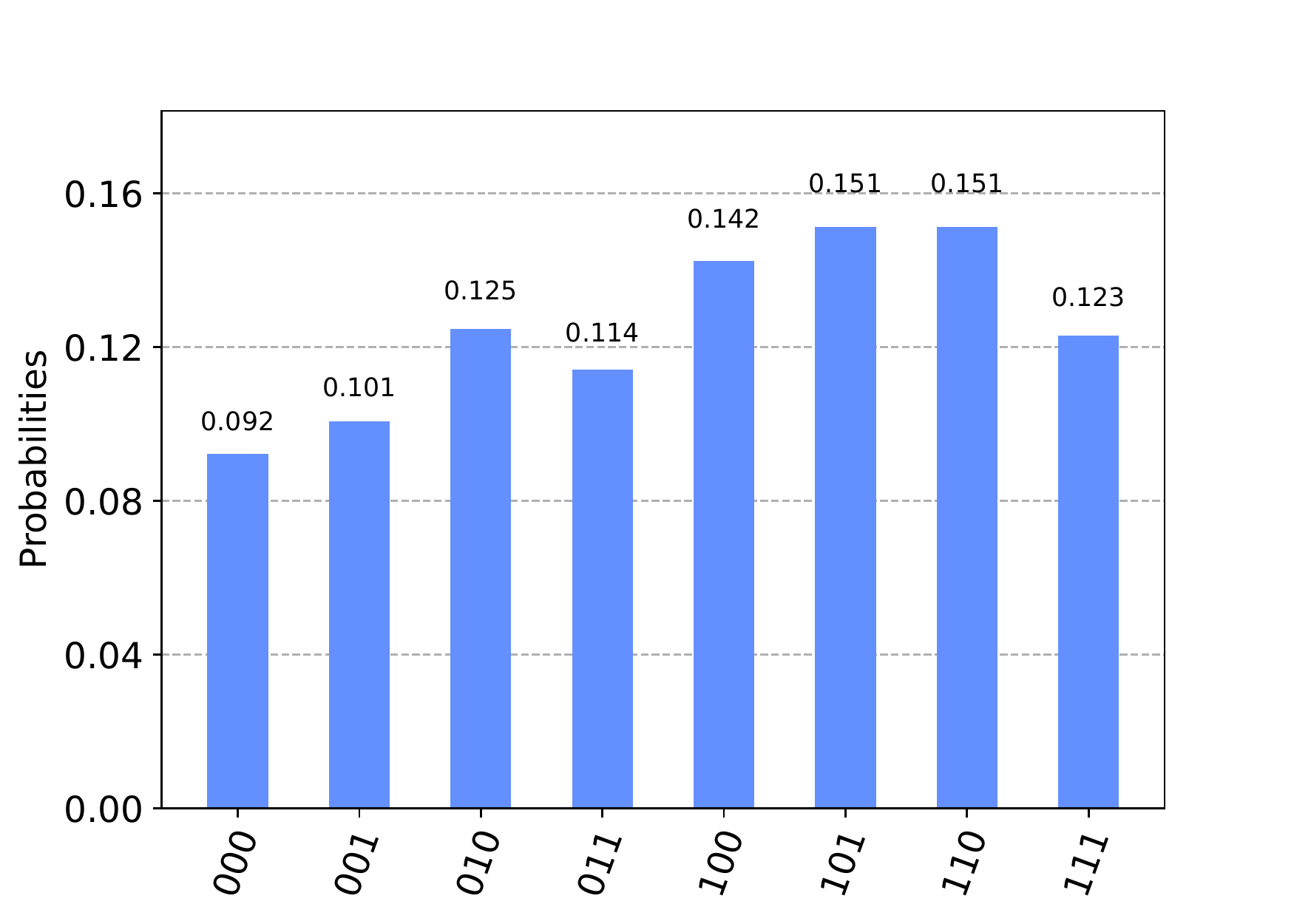}
        \caption{Actual ibmq\_lima}
        \label{fig:lima_actual}
    \end{subfigure}
    \caption{Results of Grover search. The expected outputs are 101 and 110. Probabilities are computed based on $1024$ shots for simulators and $4000$ shots for a real QC ibmq\_lima (the term \emph{shot} refers to a single experiment run). As expected, simulator of an ideal QC (Figure~\ref{fig:sim}) yields the most accurate results. Simulated ibmq\_lima results (Figure~\ref{fig:lima_sim}) are more optimistic than actual ibmq\_lima results (Figure~\ref{fig:lima_actual}). This can be explained by the actual QC's performance fluctuating due to environmental factors.}
    \label{fig:grover_result}
\end{figure*}

The output of Grover's algorithm can be seen in Figure~\ref{fig:grover_result}. According to the simulators and the real QC, 110 and 101 are the top-two returned values of $\omega$. However, due to noise, the $p$-values on the real QC are lower than those on the simulators.

Step~\ref{alg:bug_finder_step4} of Algorithm~\ref{alg:bug_finder} converts $\omega$ back to $\vec{z}$ values:
\begin{equation}\label{eq:conversion}
\begin{split}
    \underbrace{110}_{\omega}  \xrightarrow{\textrm{split}} \underbrace{11}_{z_1} \underbrace{0}_{z_2} \xrightarrow{\textrm{from binary}} \underbrace{3}_{z_1} \underbrace{\textrm{false}}_{z_2}, \\
    \underbrace{101}_{\omega}  \xrightarrow{\textrm{split}} \underbrace{10}_{z_1} \underbrace{1}_{z_2} \xrightarrow{\textrm{from binary}} \underbrace{2}_{z_1} \underbrace{\textrm{true}}_{z_2}.
\end{split}
\end{equation}
These are the values we expected. In other words, Algorithm~\ref{alg:bug_finder} correctly identified two inputs that caused errors in software under test.

\subsubsection{Computational complexity}\label{sec:ex_search_complexity}
How laborious is Algorithm~\ref{alg:bug_finder}? 

In Step~\ref{alg:bug_finder_step1}, code written in a high-level language (e.g., C/C++ or Python) is compiled into a quantum circuit. Depending on the compiler, the complexity of this process may vary, but compiler designers are aiming for $O(\alpha^2)$, where $\alpha$ is the size of the input~\cite{cooper2011engineering}. In practice, even for large code bases with millions of lines of code, compilation on modern CC is relatively fast.

The computation complexity of Steps~\ref{alg:bug_finder_step2} and \ref{alg:bug_finder_step3} is $O(\varepsilon^{-1} \sqrt{N/K})$ and $O(\sqrt{N/K})$, respectively~\cite{boyer1998tight, brassard1998quantum}. Complexity of Step~\ref{alg:bug_finder_step4} is proportional to the number of bits required to represent $x$, i.e., $O(\log_2{N})$.
Therefore, we estimate that the majority of time will be spent on Steps~\ref{alg:bug_finder_step2} and \ref{alg:bug_finder_step3}, with the dominant term $O(\varepsilon^{-1} \sqrt{N/K})$.

When comparing exhaustive testing on CC and QC, QC is faster because exhaustive search on a CC requires $O(N) = O(2^n)$ computations, while on a QC~--- only $O(\varepsilon^{-1} \sqrt{N/K}) = O(\varepsilon^{-1}2^{n/2}K^{-1/2})$ computations.

$2^{n/2}$ grows much slower\footnote{For example, suppose $n = 64$ bits. Then $2^n = 2^{64} \approx 2 \times 10^{19}$, while $2^{n/2} = 2^{32} \approx 4 \times 10^9$.} than $2^n$, but it still increases exponentially. Thus, if our software has input parameters requiring many bits, the search will become prohibitively expensive, even on a QC. Therefore, let us explore non-exhaustive testing methods.

\subsection{Non-exhaustive testing}
As discussed in Section~\ref{sec:intro}, in the non-exhaustive case, we test a subset of input parameters  $N$. Essentially, we want to reduce  $N$ and speed up testing.

An example of such an approach is the use of a $t$-way combinatorial testing to explore different combinations of parameters~\cite{kuhn2004software}. A $1$-way or $2$-way test can catch most defects, but $t > 2$ is needed to expose some~\cite{kuhn2004software}. As $t$ increases, testing costs quickly rise. Nevertheless, it may be necessary to find these defects (e.g., in mission-critical software~\cite{hynninen2018software}). 

When $t$ is large, we can use Algorithm~\ref{alg:bug_finder} to accelerate $t$-way testing. Step~\ref{alg:bug_finder_step1} of Algorithm~\ref{alg:bug_finder} needs to be modified to map the values of $\vec{z}$ to $x$. For example, a mapping function can be created by automatically generating a simple lookup table that will be encoded in the circuit.

\subsubsection{Example}
Let us consider the function $g(z_3)$, where  $z_3$ is a $64$-bit integer, i.e., $n = 64$). There are only three input values we want to test, e.g., $z_3 = \{7, 200, 1005 \}$.

We will refer to the number of inputs in a subset as $N_s$. In our example $N_s=3$. To represent the input, we only need $\ceil{\log_2{N_s}} = \ceil{\log_2{3}} = 2$ qubits rather than $64$ qubits. During compilation, we would add a data structure that mapped each of these three values to their 64-bit representations (note that we will still need $64$ auxiliary qubits to build oracle's circuit, but we will only be measuring the two input qubits).

\subsubsection{Computational complexity}\label{sec:non_ex_search_complexity}
We will get $O(\varepsilon^{-1} \sqrt{N_s/K}) = O(\varepsilon^{-1}2^{n_s/2}K^{-1/2})$ computations, where $N_s = 2^{n_s}$, if we apply the complexity analysis logic discussed in Section~\ref{sec:ex_search_complexity}. Therefore, as long as $N \gg N_s$, the non-exhaustive test will be significantly faster than the exhaustive test. Moreover, non-exhaustive testing on a CC would require $O(N_s) = O(2^{n_s})$ computations, so QC will outperform CC.

As we need to add the code to map $\vec{z}$ to $x$, the cost of executing the oracle increases compared to the exhaustive testing case. Nevertheless, such look-up tables can be constructed efficiently; e.g., a hash table's search complexity is $O(1)$. As a result, the computation time increase will be minimal.

\section{Discussion}\label{sec:limitations}
According to Section~\ref{sec:testing}, dynamic testing might be faster on QC than on CC. Does this approach have any drawbacks?
Section~\ref{sec:blockers} discusses blocking issues that prevent the algorithm from becoming practical. 
Section~\ref{sec:limiters} describes the limitations of this approach, including the types of programs that can be tested.
Section~\ref{sec:tuning} concludes with notes on tuning implementation of Algorithm~\ref{alg:bug_finder}.

\subsection{Blocking issues}\label{sec:blockers}

\paragraph{Creation of the oracle $f(x)$}
The conversion of classical circuits to quantum circuits has been an active research area for at least two decades (see~\cite{seidel2021automatic} for a review). Yet, more work must be done before practitioners can use them to convert arbitrary CC programs. Specifically, it will be difficult for practitioners to use our dynamic testing method without an efficient universal compiler. This goal can be accomplished with the help of academics and practitioners in the compiler community.

What if the time it takes to generate and compute Grover's algorithm negates the savings from the search itself? This may happen if we simply load a database into a QC~\cite{viamontes2005quantum}. The testing method presented in this paper, however, does not explicitly search for database elements but instead computes the inverse of $f(x) = 1$, thereby reducing the possibility of such an outcome~\cite{viamontes2005quantum}.

\paragraph{Number of shots}
A noisy intermediate-scale quantum (NISQ) computer requires a high number of shots to combat noise. Figure~\ref{fig:grover_result} illustrates the detrimental effects of noise (compare $p$-values of correct answers on simulators and a real QC).

A fault-tolerant QC (FTQC)~\cite{nielsen_chuang_2010}, which employs error-correcting schemes to operate on logical/ideal qubits (constructed from groups of physical qubits~\cite{chao2018fault}), will require fewer shots than NISQ device. For example, theoretically, if $K = 1$, the correct answer will likely be obtained with one shot.

However, as $K$ increases, so does the number of shots (as every shot yields a single value of $\omega$). The number of shots on FTQC can be estimated by converting our problem to a classical Coupon Collector problem~\cite{ferrante2014coupon}, which tries to determine how many coupons $S$ one needs to buy to collect $K$ distinct coupons (see~\cite{ferrante2014coupon} for review). In our case, to capture $K$ distinct inputs causing errors, we need to take, on average, $S$ shots. Assuming that the probability of observing a particular error-inducing input is $1/K$, the expected value of $S$ is given by
\begin{equation}\label{eq:expected_shots}
    \mathbb{E}[S] = K \sum_{i=1}^{K}{\frac{1}{i}} = K \ln(K) + K \gamma + \frac{1}{2} + O\left(\frac{1}{K}\right), K \to +\infty,
\end{equation}
where $\gamma \approx 0.5772$ is the Euler-Mascheroni constant (see~\cite[p. 5]{ferrante2014coupon} for details).

\paragraph{Speed of quantum computers}
In order for the oracle $f(x)$ to be executed quickly, the QC should be able to run the circuit efficiently. Modern QCs are still relatively slow: e.g., currently, top IBM's QC can process $1419$ Circuit Layer Operations Per Second (CLOPS)~\cite{wack2021scale}. A future generation of QCs will resolve this engineering problem.

\paragraph{Timeline}
Section~\ref{sec:example} demonstrates that dynamic testing of QC may be possible even on a NISQ device. As with many other quantum algorithms applicable to software engineering tasks, we will need to wait for a device with a large number of qubits (preferably with good interconnectivity), high coherence, and high fidelity before the algorithms can be used in practice~\cite{miranskyy2022quantum}. Small FTQCs with such characteristics may appear within ten years~\cite{sevilla2020forecasting}.

\subsection{Limiting issues}\label{sec:limiters}

\paragraph{CC program types}
It is theoretically possible to convert any CC program into a QC program, as we discussed in Section~\ref{sec:cc_conversion_body}. For single-threaded, single-process software, this should be relatively straightforward.

A multi-threaded, multi-process program, however, would be more challenging to convert since it would require implementation and conversion of the thread and process manager. One would need to load the code to be tested into a virtual machine and then load this combined codebase into the quantum circuit. 

A similar problem will arise with programs implemented using the async/await pattern, as we need to convert the supporting libraries.

\paragraph{CC program structure}
Estimates of computational complexity in Sections~\ref{sec:ex_search_complexity} and~\ref{sec:non_ex_search_complexity} are based on a critical assumption made by Grover: oracle $f(x)$ ``can be evaluated in unit time'' for any value of $x$~\cite{grover1996fast}. This assumption governs the computational complexity of Steps~\ref{alg:bug_finder_step2} and~\ref{alg:bug_finder_step3} of Algorithm~\ref{alg:bug_finder}.

However, this assumption is not always valid. For example, $f(x)$ may have a for-loop with $x$ iterations. Based on $\max(x) = N = 2^n$, the execution time of the Oracle, on both CC and QC, will grow exponentially with $n$. Although exhaustive testing on the QC will still be faster than on CC, the execution time may be prohibitive\footnote{Hypothetically, some for-loops may be parallelized on a QC, see~\cite[Sec. 3.5]{zomorodi2014synthesis} for review.}. Thus, we may have to create a static analyzer that checks the code of the CC program $g(\vec{z})$ before conversion and warns the user if such pathological cases are found.

Note that this pathological code may be less problematic for non-exhaustive testing, because we can limit the number of parameter values passed to $f(x)$.

\paragraph{Calls to external services}
In many programs, external services (e.g., a database or a web service) must be called. We would not be able to call these services from a QC unless such services are implemented within the QC (and that is not going to happen soon~\cite{miranskyy2019testing, miranskyy2021testing}). As an alternative, we can simulate the calls to external services using mock and stub objects  (see~\cite[Ch. 4]{osherove2013art} for review). The mocked and stubbed code can easily be converted to quantum circuit as part of the oracle generation process.

\subsection{Tuning of Algorithm~\ref{alg:bug_finder}}\label{sec:tuning}

\paragraph{Count of inputs causing defects}
Algorithm~\ref{alg:bug_finder} can potentially be stopped at Step~\ref{alg:bug_finder_step2} and simply return $K$ (i.e., the count of inputs causing errors). As a result, the developer can see how many inputs resulted in errors. However, with this information alone, the developer cannot reproduce the failures (unless $K=0$ or $K=N$, indicating that all tests pass or fail, respectively). Thus, we conjecture that a practitioner would usually want to run all the steps of Algorithm~\ref{alg:bug_finder} and obtain the actual values of inputs that cause errors.

\paragraph{Parallelism}
Grover's search in Step~\ref{alg:bug_finder_step3} of Algorithm~\ref{alg:bug_finder} will need to be repeated multiple times (especially for large values of $K$, see Eq.~\ref{eq:expected_shots}), since each call will return only one $\omega$ value (error-generating input). We can parallelize Step~\ref{alg:bug_finder_step3} (since each shot is independent) by compiling Grover's circuit and running it on multiple QCs.

\section{Summary}\label{sec:summary}
This paper showed that dynamic testing of programs written for a CC can be sped up on a QC. Specifically, computational complexity decreases from $O(N)$ to  $O(\varepsilon^{-1} \sqrt{N/K})$. Simulated and real QC outputs from a toy example were used to illustrate this approach. A discussion on the limitations of dynamically testing CC programs on a QC was conducted.

Two key issues must be solved to make the approach practical:
\begin{enumerate*}[label={(\arabic*)}]
    \item converters from classical to quantum circuits need to be improved, and
    \item QCs need to become faster, larger, and more ``robust.''
\end{enumerate*}
These two issues can be resolved, but a significant effort will be required from the academic and industrial communities.

This work may serve as a starting point for exploring the use of QC for dynamic testing of CC code. Academics and practitioners are invited to explore this promising field of research further.

\section*{Acknowledgement}
I would like to thank anonymous reviewers for their thoughtful comments and suggestions! 
This research is funded in part by NSERC Discovery Grant No. RGPIN-2022-03886.

\bibliography{references}


\begin{thebibliography}{34}


\ifx \showCODEN    \undefined \def \showCODEN     #1{\unskip}     \fi
\ifx \showDOI      \undefined \def \showDOI       #1{#1}\fi
\ifx \showISBNx    \undefined \def \showISBNx     #1{\unskip}     \fi
\ifx \showISBNxiii \undefined \def \showISBNxiii  #1{\unskip}     \fi
\ifx \showISSN     \undefined \def \showISSN      #1{\unskip}     \fi
\ifx \showLCCN     \undefined \def \showLCCN      #1{\unskip}     \fi
\ifx \shownote     \undefined \def \shownote      #1{#1}          \fi
\ifx \showarticletitle \undefined \def \showarticletitle #1{#1}   \fi
\ifx \showURL      \undefined \def \showURL       {\relax}        \fi
\providecommand\bibfield[2]{#2}
\providecommand\bibinfo[2]{#2}
\providecommand\natexlab[1]{#1}
\providecommand\showeprint[2][]{arXiv:#2}

\bibitem[\protect\citeauthoryear{??}{ibm}{2022}]%
        {ibm_quantum}
 \bibinfo{year}{2022}\natexlab{}.
\newblock \bibinfo{booktitle}{\emph{IBM Quantum}}.
\newblock
\urldef\tempurl%
\url{https://quantum-computing.ibm.com}
\showURL{%
\tempurl}


\bibitem[\protect\citeauthoryear{Aaronson}{Aaronson}{2005}]%
        {aaronson2005quantum}
\bibfield{author}{\bibinfo{person}{Scott Aaronson}.}
  \bibinfo{year}{2005}\natexlab{}.
\newblock \showarticletitle{Quantum computing, postselection, and probabilistic
  polynomial-time}.
\newblock \bibinfo{journal}{\emph{Proceedings of the Royal Society A:
  Mathematical, Physical and Engineering Sciences}} \bibinfo{volume}{461},
  \bibinfo{number}{2063} (\bibinfo{year}{2005}), \bibinfo{pages}{3473--3482}.
\newblock


\bibitem[\protect\citeauthoryear{Aaronson and Rall}{Aaronson and Rall}{2020}]%
        {aaronson2020quantum}
\bibfield{author}{\bibinfo{person}{Scott Aaronson} {and}
  \bibinfo{person}{Patrick Rall}.} \bibinfo{year}{2020}\natexlab{}.
\newblock \showarticletitle{Quantum approximate counting, simplified}. In
  \bibinfo{booktitle}{\emph{Symposium on Simplicity in Algorithms}}. SIAM,
  \bibinfo{pages}{24--32}.
\newblock


\bibitem[\protect\citeauthoryear{Abbas, Andersson, Asfaw, Corcoles, Bello,
  Ben-Haim, Bozzo-Rey, Bravyi, Bronn, Capelluto, Vazquez, Ceroni, Chen, Frisch,
  Gambetta, Garion, Gil, Gonzalez, Harkins, Imamichi, Jayasinha, Kang,
  h.~Karamlou, Loredo, McKay, Maldonado, Macaluso, Mezzacapo, Minev, Movassagh,
  Nannicini, Nation, Phan, Pistoia, Rattew, Schaefer, Shabani, Smolin, Stenger,
  Temme, Tod, Wanzambi, Wood, and Wootton.}{Abbas et~al\mbox{.}}{2020}]%
        {Qiskit-Textbook}
\bibfield{author}{\bibinfo{person}{Amira Abbas}, \bibinfo{person}{Stina
  Andersson}, \bibinfo{person}{Abraham Asfaw}, \bibinfo{person}{Antonio
  Corcoles}, \bibinfo{person}{Luciano Bello}, \bibinfo{person}{Yael Ben-Haim},
  \bibinfo{person}{Mehdi Bozzo-Rey}, \bibinfo{person}{Sergey Bravyi},
  \bibinfo{person}{Nicholas Bronn}, \bibinfo{person}{Lauren Capelluto},
  \bibinfo{person}{Almudena~Carrera Vazquez}, \bibinfo{person}{Jack Ceroni},
  \bibinfo{person}{Richard Chen}, \bibinfo{person}{Albert Frisch},
  \bibinfo{person}{Jay Gambetta}, \bibinfo{person}{Shelly Garion},
  \bibinfo{person}{Leron Gil}, \bibinfo{person}{Salvador De La~Puente
  Gonzalez}, \bibinfo{person}{Francis Harkins}, \bibinfo{person}{Takashi
  Imamichi}, \bibinfo{person}{Pavan Jayasinha}, \bibinfo{person}{Hwajung Kang},
  \bibinfo{person}{Amir h. Karamlou}, \bibinfo{person}{Robert Loredo},
  \bibinfo{person}{David McKay}, \bibinfo{person}{Alberto Maldonado},
  \bibinfo{person}{Antonio Macaluso}, \bibinfo{person}{Antonio Mezzacapo},
  \bibinfo{person}{Zlatko Minev}, \bibinfo{person}{Ramis Movassagh},
  \bibinfo{person}{Giacomo Nannicini}, \bibinfo{person}{Paul Nation},
  \bibinfo{person}{Anna Phan}, \bibinfo{person}{Marco Pistoia},
  \bibinfo{person}{Arthur Rattew}, \bibinfo{person}{Joachim Schaefer},
  \bibinfo{person}{Javad Shabani}, \bibinfo{person}{John Smolin},
  \bibinfo{person}{John Stenger}, \bibinfo{person}{Kristan Temme},
  \bibinfo{person}{Madeleine Tod}, \bibinfo{person}{Ellinor Wanzambi},
  \bibinfo{person}{Stephen Wood}, {and} \bibinfo{person}{James Wootton.}}
  \bibinfo{year}{2020}\natexlab{}.
\newblock \bibinfo{title}{Learn Quantum Computation Using Qiskit}.
\newblock
\newblock
\urldef\tempurl%
\url{http://community.qiskit.org/textbook}
\showURL{%
\tempurl}


\bibitem[\protect\citeauthoryear{Amy and Gheorghiu}{Amy and Gheorghiu}{2020}]%
        {amy2020staq}
\bibfield{author}{\bibinfo{person}{Matthew Amy} {and} \bibinfo{person}{Vlad
  Gheorghiu}.} \bibinfo{year}{2020}\natexlab{}.
\newblock \showarticletitle{staq—A full-stack quantum processing toolkit}.
\newblock \bibinfo{journal}{\emph{Quantum Science and Technology}}
  \bibinfo{volume}{5}, \bibinfo{number}{3} (\bibinfo{year}{2020}),
  \bibinfo{pages}{034016}.
\newblock


\bibitem[\protect\citeauthoryear{Benioff}{Benioff}{1980}]%
        {benioff1980computer}
\bibfield{author}{\bibinfo{person}{Paul Benioff}.}
  \bibinfo{year}{1980}\natexlab{}.
\newblock \showarticletitle{The computer as a physical system: A microscopic
  quantum mechanical Hamiltonian model of computers as represented by Turing
  machines}.
\newblock \bibinfo{journal}{\emph{Journal of statistical physics}}
  \bibinfo{volume}{22}, \bibinfo{number}{5} (\bibinfo{year}{1980}),
  \bibinfo{pages}{563--591}.
\newblock


\bibitem[\protect\citeauthoryear{Bennett, Bernstein, Brassard, and
  Vazirani}{Bennett et~al\mbox{.}}{1997}]%
        {bennett1997strengths}
\bibfield{author}{\bibinfo{person}{Charles~H Bennett}, \bibinfo{person}{Ethan
  Bernstein}, \bibinfo{person}{Gilles Brassard}, {and} \bibinfo{person}{Umesh
  Vazirani}.} \bibinfo{year}{1997}\natexlab{}.
\newblock \showarticletitle{Strengths and weaknesses of quantum computing}.
\newblock \bibinfo{journal}{\emph{SIAM journal on Computing}}
  \bibinfo{volume}{26}, \bibinfo{number}{5} (\bibinfo{year}{1997}),
  \bibinfo{pages}{1510--1523}.
\newblock


\bibitem[\protect\citeauthoryear{Bojić}{Bojić}{2014}]%
        {bojic2014approach}
\bibfield{author}{\bibinfo{person}{Alan Bojić}.}
  \bibinfo{year}{2014}\natexlab{}.
\newblock \showarticletitle{An approach to source code conversion of classical
  programming languages into source code of quantum programming languages}.
\newblock \bibinfo{journal}{\emph{Journal of Information and Organizational
  Sciences}} \bibinfo{volume}{38}, \bibinfo{number}{2} (\bibinfo{year}{2014}),
  \bibinfo{pages}{75--82}.
\newblock


\bibitem[\protect\citeauthoryear{Boyer, Brassard, H{\o}yer, and Tapp}{Boyer
  et~al\mbox{.}}{1998}]%
        {boyer1998tight}
\bibfield{author}{\bibinfo{person}{Michel Boyer}, \bibinfo{person}{Gilles
  Brassard}, \bibinfo{person}{Peter H{\o}yer}, {and} \bibinfo{person}{Alain
  Tapp}.} \bibinfo{year}{1998}\natexlab{}.
\newblock \showarticletitle{Tight bounds on quantum searching}.
\newblock \bibinfo{journal}{\emph{Fortschritte der Physik: Progress of
  Physics}} \bibinfo{volume}{46}, \bibinfo{number}{4-5} (\bibinfo{year}{1998}),
  \bibinfo{pages}{493--505}.
\newblock


\bibitem[\protect\citeauthoryear{Brassard, H{\o}yer, and Tapp}{Brassard
  et~al\mbox{.}}{1998}]%
        {brassard1998quantum}
\bibfield{author}{\bibinfo{person}{Gilles Brassard}, \bibinfo{person}{Peter
  H{\o}yer}, {and} \bibinfo{person}{Alain Tapp}.}
  \bibinfo{year}{1998}\natexlab{}.
\newblock \showarticletitle{Quantum counting}. In
  \bibinfo{booktitle}{\emph{International Colloquium on Automata, Languages,
  and Programming}}. Springer, \bibinfo{pages}{820--831}.
\newblock


\bibitem[\protect\citeauthoryear{Chao and Reichardt}{Chao and
  Reichardt}{2018}]%
        {chao2018fault}
\bibfield{author}{\bibinfo{person}{Rui Chao} {and} \bibinfo{person}{Ben~W
  Reichardt}.} \bibinfo{year}{2018}\natexlab{}.
\newblock \showarticletitle{Fault-tolerant quantum computation with few
  qubits}.
\newblock \bibinfo{journal}{\emph{npj Quantum Information}}
  \bibinfo{volume}{4}, \bibinfo{number}{1} (\bibinfo{year}{2018}),
  \bibinfo{pages}{1--8}.
\newblock


\bibitem[\protect\citeauthoryear{Cooper and Torczon}{Cooper and
  Torczon}{2011}]%
        {cooper2011engineering}
\bibfield{author}{\bibinfo{person}{Keith~D Cooper} {and} \bibinfo{person}{Linda
  Torczon}.} \bibinfo{year}{2011}\natexlab{}.
\newblock \bibinfo{booktitle}{\emph{Engineering a compiler}
  (\bibinfo{edition}{2} ed.)}.
\newblock \bibinfo{publisher}{Elsevier}.
\newblock


\bibitem[\protect\citeauthoryear{Ferrante and Saltalamacchia}{Ferrante and
  Saltalamacchia}{2014}]%
        {ferrante2014coupon}
\bibfield{author}{\bibinfo{person}{Marco Ferrante} {and}
  \bibinfo{person}{Monica Saltalamacchia}.} \bibinfo{year}{2014}\natexlab{}.
\newblock \showarticletitle{The coupon collector's problem}.
\newblock \bibinfo{journal}{\emph{Materials matem{\`a}tics}}
  (\bibinfo{year}{2014}), \bibinfo{pages}{1--35}.
\newblock


\bibitem[\protect\citeauthoryear{Grover}{Grover}{1996}]%
        {grover1996fast}
\bibfield{author}{\bibinfo{person}{Lov~K Grover}.}
  \bibinfo{year}{1996}\natexlab{}.
\newblock \showarticletitle{A fast quantum mechanical algorithm for database
  search}. In \bibinfo{booktitle}{\emph{28th annual ACM symposium on theory of
  computing}}. ACM, \bibinfo{pages}{212--219}.
\newblock


\bibitem[\protect\citeauthoryear{Holmes}{Holmes}{2016}]%
        {holmes2016rkqc}
\bibfield{author}{\bibinfo{person}{Adam Holmes}.}
  \bibinfo{year}{2016}\natexlab{}.
\newblock \bibinfo{booktitle}{\emph{epiqc/RKQC: RKQC is a compiler for
  reversible logic circuitry}}.
\newblock
\urldef\tempurl%
\url{https://github.com/epiqc/RKQC}
\showURL{%
\tempurl}


\bibitem[\protect\citeauthoryear{Hynninen, Kasurinen, Knutas, and
  Taipale}{Hynninen et~al\mbox{.}}{2018}]%
        {hynninen2018software}
\bibfield{author}{\bibinfo{person}{Timo Hynninen}, \bibinfo{person}{Jussi
  Kasurinen}, \bibinfo{person}{Antti Knutas}, {and} \bibinfo{person}{Ossi
  Taipale}.} \bibinfo{year}{2018}\natexlab{}.
\newblock \showarticletitle{Software testing: Survey of the industry
  practices}. In \bibinfo{booktitle}{\emph{41st International Convention on
  Information and Communication Technology, Electronics and Microelectronics
  (MIPRO)}}. IEEE, \bibinfo{pages}{1449--1454}.
\newblock


\bibitem[\protect\citeauthoryear{JavadiAbhari, Patil, Kudrow, Heckey, Lvov,
  Chong, and Martonosi}{JavadiAbhari et~al\mbox{.}}{2014}]%
        {javadiabhari2014scaffcc}
\bibfield{author}{\bibinfo{person}{Ali JavadiAbhari}, \bibinfo{person}{Shruti
  Patil}, \bibinfo{person}{Daniel Kudrow}, \bibinfo{person}{Jeff Heckey},
  \bibinfo{person}{Alexey Lvov}, \bibinfo{person}{Frederic~T Chong}, {and}
  \bibinfo{person}{Margaret Martonosi}.} \bibinfo{year}{2014}\natexlab{}.
\newblock \showarticletitle{ScaffCC: A framework for compilation and analysis
  of quantum computing programs}. In \bibinfo{booktitle}{\emph{11th ACM
  Conference on Computing Frontiers}}. \bibinfo{publisher}{ACM},
  \bibinfo{pages}{1--10}.
\newblock


\bibitem[\protect\citeauthoryear{Kuhn, Wallace, and Gallo}{Kuhn
  et~al\mbox{.}}{2004}]%
        {kuhn2004software}
\bibfield{author}{\bibinfo{person}{D~Richard Kuhn}, \bibinfo{person}{Dolores~R
  Wallace}, {and} \bibinfo{person}{Albert~M Gallo}.}
  \bibinfo{year}{2004}\natexlab{}.
\newblock \showarticletitle{Software fault interactions and implications for
  software testing}.
\newblock \bibinfo{journal}{\emph{IEEE trans. on softw. eng.}}
  \bibinfo{volume}{30}, \bibinfo{number}{6} (\bibinfo{year}{2004}),
  \bibinfo{pages}{418--421}.
\newblock


\bibitem[\protect\citeauthoryear{Miller, Maslov, and Dueck}{Miller
  et~al\mbox{.}}{2003}]%
        {miller2003transformation}
\bibfield{author}{\bibinfo{person}{D~Michael Miller}, \bibinfo{person}{Dmitri
  Maslov}, {and} \bibinfo{person}{Gerhard~W Dueck}.}
  \bibinfo{year}{2003}\natexlab{}.
\newblock \showarticletitle{A transformation based algorithm for reversible
  logic synthesis}. In \bibinfo{booktitle}{\emph{Design automation
  conference}}. IEEE, \bibinfo{pages}{318--323}.
\newblock


\bibitem[\protect\citeauthoryear{Miranskyy}{Miranskyy}{2022}]%
        {andriy_miranskyy_2022_7065888}
\bibfield{author}{\bibinfo{person}{Andriy Miranskyy}.}
  \bibinfo{year}{2022}\natexlab{}.
\newblock \bibinfo{booktitle}{\emph{{Supplementary material: code listing}}}.
\newblock
\urldef\tempurl%
\url{https://doi.org/10.5281/zenodo.7065888}
\showDOI{\tempurl}


\bibitem[\protect\citeauthoryear{Miranskyy, Khan, Faye, and Mendes}{Miranskyy
  et~al\mbox{.}}{2022}]%
        {miranskyy2022quantum}
\bibfield{author}{\bibinfo{person}{Andriy Miranskyy}, \bibinfo{person}{Mushahid
  Khan}, \bibinfo{person}{Jean Paul~Latyr Faye}, {and} \bibinfo{person}{Udson~C
  Mendes}.} \bibinfo{year}{2022}\natexlab{}.
\newblock \showarticletitle{Quantum Computing for Software Engineering:
  Prospects}.
\newblock \bibinfo{journal}{\emph{arXiv preprint arXiv:2203.03575}}
  (\bibinfo{year}{2022}).
\newblock
\newblock
\shownote{To appear in Proceedings of the 1st International Workshop on Quantum
  Programming for Software Engineering.}


\bibitem[\protect\citeauthoryear{Miranskyy and Zhang}{Miranskyy and
  Zhang}{2019}]%
        {miranskyy2019testing}
\bibfield{author}{\bibinfo{person}{Andriy Miranskyy} {and} \bibinfo{person}{Lei
  Zhang}.} \bibinfo{year}{2019}\natexlab{}.
\newblock \showarticletitle{On testing quantum programs}. In
  \bibinfo{booktitle}{\emph{41st Int. Conf. on Softw. Eng.: New Ideas and
  Emerging Results (ICSE-NIER)}}. IEEE, \bibinfo{pages}{57--60}.
\newblock


\bibitem[\protect\citeauthoryear{Miranskyy, Zhang, and Doliskani}{Miranskyy
  et~al\mbox{.}}{2021}]%
        {miranskyy2021testing}
\bibfield{author}{\bibinfo{person}{Andriy Miranskyy}, \bibinfo{person}{Lei
  Zhang}, {and} \bibinfo{person}{Javad Doliskani}.}
  \bibinfo{year}{2021}\natexlab{}.
\newblock \showarticletitle{On testing and debugging quantum software}.
\newblock \bibinfo{journal}{\emph{arXiv preprint arXiv:2103.09172}}
  (\bibinfo{year}{2021}).
\newblock


\bibitem[\protect\citeauthoryear{Nielsen and Chuang}{Nielsen and
  Chuang}{2010}]%
        {nielsen_chuang_2010}
\bibfield{author}{\bibinfo{person}{Michael~A. Nielsen} {and}
  \bibinfo{person}{Isaac~L. Chuang}.} \bibinfo{year}{2010}\natexlab{}.
\newblock \bibinfo{booktitle}{\emph{Quantum Computation and Quantum
  Information: 10th Anniversary Edition}}.
\newblock \bibinfo{publisher}{Cambridge Univ. Press}.
\newblock


\bibitem[\protect\citeauthoryear{Osherove}{Osherove}{2013}]%
        {osherove2013art}
\bibfield{author}{\bibinfo{person}{Roy Osherove}.}
  \bibinfo{year}{2013}\natexlab{}.
\newblock \bibinfo{booktitle}{\emph{The Art of Unit Testing: with examples in
  C\#} (\bibinfo{edition}{2} ed.)}.
\newblock \bibinfo{publisher}{Manning}.
\newblock


\bibitem[\protect\citeauthoryear{Pressman and Maxim}{Pressman and
  Maxim}{2015}]%
        {pressman2005software}
\bibfield{author}{\bibinfo{person}{Roger~S Pressman} {and}
  \bibinfo{person}{Bruce~R. Maxim}.} \bibinfo{year}{2015}\natexlab{}.
\newblock \bibinfo{booktitle}{\emph{Software engineering: a practitioner's
  approach} (\bibinfo{edition}{8} ed.)}.
\newblock \bibinfo{publisher}{McGraw Hill}.
\newblock


\bibitem[\protect\citeauthoryear{Schmitt and De~Micheli}{Schmitt and
  De~Micheli}{2022}]%
        {schmitt2022tweedledum}
\bibfield{author}{\bibinfo{person}{Bruno Schmitt} {and}
  \bibinfo{person}{Giovanni De~Micheli}.} \bibinfo{year}{2022}\natexlab{}.
\newblock \showarticletitle{tweedledum: a compiler companion for quantum
  computing}. In \bibinfo{booktitle}{\emph{2022 Design, Automation \& Test in
  Europe Conference \& Exhibition (DATE)}}. IEEE, \bibinfo{pages}{7--12}.
\newblock


\bibitem[\protect\citeauthoryear{Seidel, Becker, Bock, Tcholtchev, Gheorge-Pop,
  and Hauswirth}{Seidel et~al\mbox{.}}{2021}]%
        {seidel2021automatic}
\bibfield{author}{\bibinfo{person}{Raphael Seidel}, \bibinfo{person}{Colin
  Kai-Uwe Becker}, \bibinfo{person}{Sebastian Bock}, \bibinfo{person}{Nikolay
  Tcholtchev}, \bibinfo{person}{Ilie-Daniel Gheorge-Pop}, {and}
  \bibinfo{person}{Manfred Hauswirth}.} \bibinfo{year}{2021}\natexlab{}.
\newblock \showarticletitle{Automatic Generation of Grover Quantum Oracles for
  Arbitrary Data Structures}.
\newblock \bibinfo{journal}{\emph{arXiv preprint arXiv:2110.07545}}
  (\bibinfo{year}{2021}).
\newblock


\bibitem[\protect\citeauthoryear{Sevilla and Riedel}{Sevilla and
  Riedel}{2020}]%
        {sevilla2020forecasting}
\bibfield{author}{\bibinfo{person}{Jaime Sevilla} {and} \bibinfo{person}{C~Jess
  Riedel}.} \bibinfo{year}{2020}\natexlab{}.
\newblock \showarticletitle{Forecasting timelines of quantum computing}.
\newblock \bibinfo{journal}{\emph{arXiv preprint arXiv:2009.05045}}
  (\bibinfo{year}{2020}).
\newblock


\bibitem[\protect\citeauthoryear{Shor}{Shor}{1994}]%
        {shor1994algorithms}
\bibfield{author}{\bibinfo{person}{Peter~W Shor}.}
  \bibinfo{year}{1994}\natexlab{}.
\newblock \showarticletitle{Algorithms for quantum computation: discrete
  logarithms and factoring}. In \bibinfo{booktitle}{\emph{35th annual symp. on
  foundations of comp. sci.}} IEEE, \bibinfo{pages}{124--134}.
\newblock


\bibitem[\protect\citeauthoryear{Viamontes, Markov, and Hayes}{Viamontes
  et~al\mbox{.}}{2005}]%
        {viamontes2005quantum}
\bibfield{author}{\bibinfo{person}{George~F Viamontes}, \bibinfo{person}{Igor~L
  Markov}, {and} \bibinfo{person}{John~P Hayes}.}
  \bibinfo{year}{2005}\natexlab{}.
\newblock \showarticletitle{Is quantum search practical?}
\newblock \bibinfo{journal}{\emph{Computing in science \& engineering}}
  \bibinfo{volume}{7}, \bibinfo{number}{3} (\bibinfo{year}{2005}),
  \bibinfo{pages}{62--70}.
\newblock


\bibitem[\protect\citeauthoryear{Wack, Paik, Javadi-Abhari, Jurcevic, Faro,
  Gambetta, and Johnson}{Wack et~al\mbox{.}}{2021}]%
        {wack2021scale}
\bibfield{author}{\bibinfo{person}{Andrew Wack}, \bibinfo{person}{Hanhee Paik},
  \bibinfo{person}{Ali Javadi-Abhari}, \bibinfo{person}{Petar Jurcevic},
  \bibinfo{person}{Ismael Faro}, \bibinfo{person}{Jay~M. Gambetta}, {and}
  \bibinfo{person}{Blake~R. Johnson}.} \bibinfo{year}{2021}\natexlab{}.
\newblock \showarticletitle{Scale, Quality, and Speed: three key attributes to
  measure the performance of near-term quantum computers}.
\newblock \bibinfo{journal}{\emph{arXiv preprint arXiv:2110.14108}}
  (\bibinfo{year}{2021}).
\newblock


\bibitem[\protect\citeauthoryear{Wang, Luo, Li, Qu, and Wang}{Wang
  et~al\mbox{.}}{2016}]%
        {wang2016improved}
\bibfield{author}{\bibinfo{person}{Feng Wang}, \bibinfo{person}{Mingxing Luo},
  \bibinfo{person}{Huiran Li}, \bibinfo{person}{Zhiguo Qu}, {and}
  \bibinfo{person}{Xiaojun Wang}.} \bibinfo{year}{2016}\natexlab{}.
\newblock \showarticletitle{Improved quantum ripple-carry addition circuit}.
\newblock \bibinfo{journal}{\emph{Science China Information Sciences}}
  \bibinfo{volume}{59}, \bibinfo{number}{4} (\bibinfo{year}{2016}),
  \bibinfo{pages}{1--8}.
\newblock


\bibitem[\protect\citeauthoryear{Zomorodi-Moghadam, Taherkhani, and
  Navi}{Zomorodi-Moghadam et~al\mbox{.}}{2014}]%
        {zomorodi2014synthesis}
\bibfield{author}{\bibinfo{person}{Mariam Zomorodi-Moghadam},
  \bibinfo{person}{Mohammad-Amin Taherkhani}, {and} \bibinfo{person}{Keivan
  Navi}.} \bibinfo{year}{2014}\natexlab{}.
\newblock \showarticletitle{Synthesis and optimization by quantum circuit
  description language}.
\newblock In \bibinfo{booktitle}{\emph{Transactions on Computational Science
  XXIV}}. \bibinfo{publisher}{Springer}, \bibinfo{pages}{74--91}.
\newblock


\end{thebibliography}
\end{document}